# Augmented smartphone bilirubinometer enabled by a mobile app that turns smartphone into multispectral imager

Qinghua He[1, 2, ‡]; Wanyu Li[3, ‡]; Yaping Shi[2]; Yi Yu[1]; Yi Zhang[2]; Wenqian Geng[3]; Zhiyuan Sun[1, *]; Ruikang K Wang[2, 4, *]

[1]Changchun Institute of Optics, Fine Mechanics and Physics, Chinese Academy of Science, Changchun, Jilin 130033, China

[2]Department of Bioengineering, University of Washington, Seattle, Washington 98105, USA

[3]Department of Hepatobiliary and pancreatic Medicine, The first Hospital of Jilin University NO.71 Xinmin Street, Changchun, Jilin 130021, China

[4]Department of Ophthalmology, University of Washington, Seattle, Washington 98109, USA

[‡]These authors contributed equally to this work.

[*]sunzhiyuan@ciomp.ac.cn

## Abstract

We present the development of SpeCamX, a mobile application that transforms any unmodified smartphone into a powerful multispectral imager capable of capturing multispectral information. Our application includes an augmented bilirubinometer, enabling accurate prediction of blood bilirubin levels (BBL). In a clinical study involving 320 patients with liver diseases, we used SpeCamX to image the bulbar conjunctiva region, and we employed a hybrid machine learning prediction model to predict BBL. We observed a high correlation with blood test results, demonstrating the efficacy of our approach. Furthermore, we compared our method, which uses spectrally augmented learning (SAL), with traditional learning based on RGB photographs (RGBL), and our results clearly indicate that SpeCamX outperforms RGBL in terms of prediction accuracy, efficiency, and stability. This study highlights the potential of SpeCamX to improve the prediction of bio-chromophores, and its ability to transform an ordinary smartphone into a powerful medical tool without the need for additional investments or expertise. This makes it suitable for widespread use, particularly in areas where medical resources are scarce.

## Funding

Washington Research Foundation

## Introduction

The smartphone steadily makes its way to become an indispensable tool in individual healthcare and living quality monitoring. This trend is made possible by the rapid developments of the sensing modules that are specifically tailored for smartphones, e.g., built-in cameras,[1-3] microphones[4-6] and touch screens.[7,8] Among these sensing modules, the camera has experienced extensive technical innovation and nowadays can deliver a comparable imaging quality to specialized medical imagers. With assembled color filters in Bayer arrangement, smartphone cameras may be able to differentiate the spectral information from collected signals in RGB channels. Utilizing this ability, researchers have been exploring smartphone-enabled analysis to extract tissue chromophores information like hemoglobin,[9] melanin,[10] bilirubin,[11] etc. However, the built-in architecture of Bayer filter in the smartphone camera inevitably limits the spectral resolution that is required to acquire/predict accurate information of the optical properties of the sample, largely due to the overlap in the sensitivity ranges in the RGB channels. [12]

For example, as a biomarker of liver functions, bilirubin has distinct absorption in the wavelength bands between 350 and 500 nm, which can be exploited to develop an optical bilirubinometer for measuring BBL in people.[13-15] Aiming for a low cost and easy access, enormous effort has been paid to realize the BBL detection with smartphone cameras. Previous studies reported some strategies by extracting raw signals from RGB channels in photographs.[16,17] While the results are promising, the measurement accuracy remains inadequate to inform clinical information. Some other studies adopted additional color calibration, image segmentation and feature extraction steps to preprocess the data to retrieve more spectral information from acquired color images.[18-20] Though the accuracy has improved, the added operations often require professional interventions that need to be accomplished off-line. In this case, the smartphone is simply used as a data collection unit for experts rather than a ready-to-use customer device, which challenges its utilities to serve the general public.

Multispectral imaging is capable of maximally recording the spectral information of subjects, thus being widely used in conducting life science research, and contributing to public healthcare services.[21-24] Realizing this technique on smartphones would create another space for exploitation to benefit our community, given the large user base, high usage frequency and low cost. A number of proof-of-concept studies were attempted where the strategy was to develop additional hardware attachments to enable the smartphone to acquire multispectral information.[25] In these attachments, diverse wavelength selection units can be designed, like light sources at appropriate wavelengths or tunable optical filters. However, asynchronous data acquisition at different wavelengths would inevitably cause a co-registration error in the data cube which consequently contaminates the analysis outcomes. More importantly, the required investments in these attachments may also pose an obstacle on their practical applications.

In this study, we describe a beta-version of mobile app, termed as SpeCamX, aimed to transform smartphones into multispectral medical imagers without any additional hardware attachments or internal modifications. With this app, we aimed to augment conventional smartphone-enabled medical imaging to realize predictions of BBL with higher quality, stability and efficiency.

**Methods**

**Study design and population**

Between September, 2021 and April, 2022, a total of 320 patients with liver diseases were enrolled in this study. Their diagnostic information was shown in **Table 1**. Anonymized and de-identified sclera images at the anterior segment of the eye were collected from the First Hospital of Jilin University (Jilin, China) using SpeCamX. The sample size was selected based on previous experience using smartphone bilirubinometer.[16-20] This value is larger than the required size derived from a power analysis so we can investigate the learning efficiency of SAL and RGBL in different sample sizes. We randomly selected the patients with liver diseases and included all the samples with accessible images of sclera regions. Blinding was applied during all data analysis. This study adhered to tenets of the Declaration of Helsinki and was performed in accordance with the Health Insurance Portability and Accountability Act. Ethical approval was obtained from the Institutional Review Board of the First Hospital of Jilin University. Written informed consent was obtained from the subject prior to the start of each study session.

Table 1: Study population

| Gender | Number |
|---|---|
| Male | 189 (59%) |
| Female | 131 (41%) |
| **Ethnicity** | **Number** |
| Asian | 320 (100%) |
| **Age** | **Number** |
| 20-29 | 10 (3%) |
| 30-39 | 32 (10%) |
| 40-49 | 92 (29%) |
| 50-59 | 94 (29%) |
| 60-69 | 69 (22%) |
| 70-79 | 19 (6%) |
| 80-89 | 4 (1%) |
| **Gender** | **Age, years** |
| Male | 52.8(12.5;21-81) |
| Female | 53.9(11.5;22-85) |
| Combined | 53.2(12.1;21-85) |

| Disorder | Number |
|---|---|
| Viral Hepatitis | 53 (17%) |
| Drug-induced liver injury | 30 (9%) |
| Alcoholic cirrhosis | 26 (8%) |
| Acute-on-chronic liver failure | 24 (7%) |
| Hepatitis B | 19 (6%) |
| Primary carcinoma of the liver | 17 (5%) |
| Acute pancreatitis | 15 (5%) |
| Drug-induced liver cirrhosis | 11 (3%) |
| Cirrhosis | 8 (3%) |
| Subacute liver failure | 7 (2%) |
| Primary biliary cirrhosis | 6 (2%) |
| Primary biliary cholangitis | 6 (2%) |
| Alcohol-induce liver injury | 6 (2%) |
| Acute liver failure | 4 (1%) |
| Other diseases | 59 (18%) |
| Multiple liver diseases | 29 (9%) |

Data are n(%) or mean (SD; range)

**Procedures**

Before imaging, the patient was asked to lie down in the bed and keep their eyes open. The operator held the smartphone that was pointing to the sclera tissue in the front eye segment. Then, the patient was asked to brink before taking each photo of the targeted region on the sclera. Ten snapshots were taken for each participant. Each photo was then reconstructed into a multispectral data cube as described before. For each snapshot, Ten ROIs were selected to calculate the averaged reflectance spectrum. Finally, the spectrum averaged from ten snapshots was used to predict BBL. The photographing was conducted in a dark ward and the subject was illuminated by the smartphone flashlight, so no further color calibration is needed. Occasionally, some trials were conducted with some residue ambient light or room light. In this case, the smartphone was recalibrated on-site with a standard color chart. Within six hours after imaging, the participants were subjected to standard clinical blood sampling to obtain their BBL. A 3mL blood sample was drawn from subcutaneous veins in the arm. Blood samples were analyzed by the diazo method using the Beckmann biochemical analysis system (Beckman Coulter Inc., CA) at the clinical pathology laboratory.

With the SpeCamX, we enable the unmodified smartphone to provide a 27-channel multispectral data cube ranging from 420 to 680 nm by a single snapshot. From the multispectral data cube, the SpeCamX provides multiple functions to estimate corresponding chromophores levels, including hemoglobin, pigmentation, bilirubin, etc. To show the performance of this app and method, we installed SpeCamX on an unmodified smartphone and used it as a bilirubinometer to quantify the sclera pigmentation at the region of bulbar conjunctiva to predict BBL (Fig. 1). In brief, the multispectral data cube of the sclera is captured to derive the reflectance spectra, which is used as the input of a hybrid prediction model trained using artificial neural network (ANN), support vector machine (SVM), k-nearest neighbors (KNN) and random forest (RF) algorithms to predict BBL. To show the benefits, we compared the prediction using SAL and RGBL.

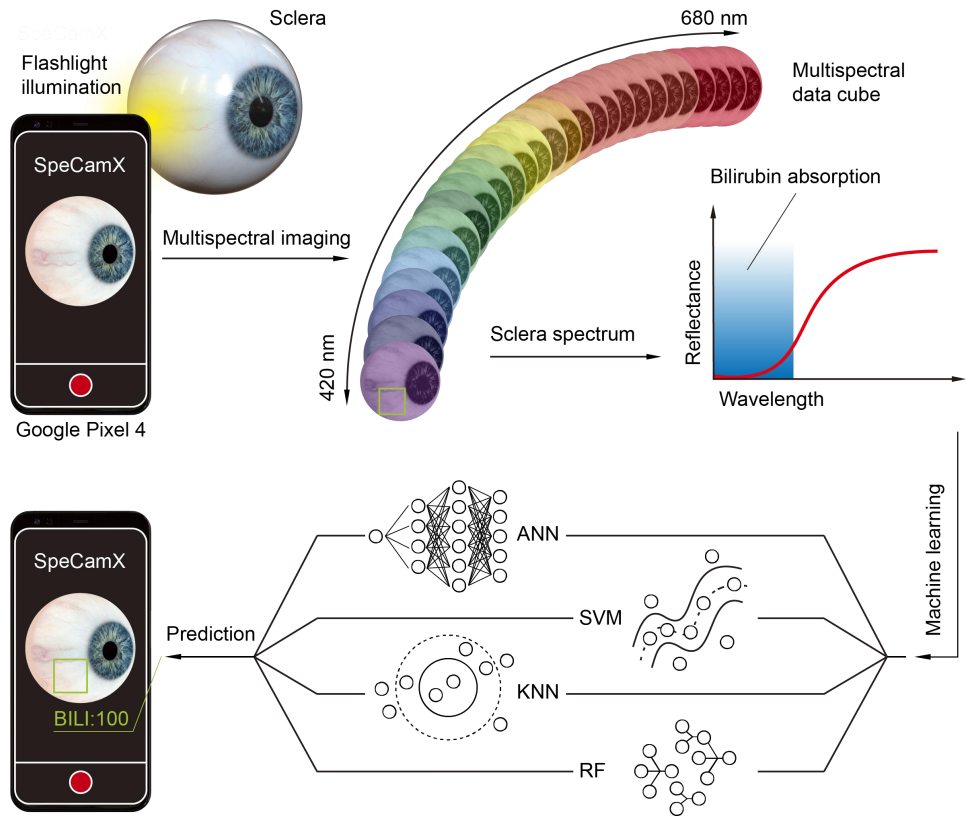

**Fig. 1 Schematic of SpeCamX-augmented smartphone bilirubinometer.** The housed flashlight in a smartphone illuminates the sclera region of the bulbar conjunctiva in a patient. The SpeCamX converts RGB photographs into multispectral data cubes ranging from 420 to 680nm that are used to derive the reflectance spectra of the sclera, which is then used to predict the BBL in patient through an incorporated regression model hybridized by four machine learning algorithms. The predicted value is immediately displayed on the smartphone screen.

**Mobile app**

In our previous study, we described a strategy to reconstruct multispectral information directly from skin photographs in RGB format.[10] Before imaging, the RGB camera was calibrated to obtain a transformation matrix using Wiener estimation algorithm. Then, using the pre-stored transformation matrix, any acquired RGB photographs can be reconstructed into multispectral images. Following this strategy, we herein optimized and expanded the workflow into SpeCamX to implement a multispectral imager with derived medical functions on an unmodified smartphone.

We calculated and embedded several transformation matrices in SpeCamX as default options for different phone models, including those from Google, Apple, Samsung and Huawei. For this purpose, we acquired RGB photographs of *X-rite ColorChecker Classic* with corresponding smartphones under the illumination provided by their flashlights. The color chart should be placed at a suitable distance with a smartphone to prevent overexposure (Maximum RGB value > 255) and under exposure (Maximum RGB value < 100). From the photographs, we sampled and calculated averaged RGB values of 24 color blocks. Then, we calculated the TM from the following equation:

$$W = <V^{'}V^t><VV^t>^{-1}$$

where $W$ is the transformation matrix. $V'$ is the reflectance spectrum of each color block.[26,27] $V$ is the RGB responses of each color block. <> is an ensemble-averaging operator, thus the matrix can be calculated from the averaged signals of 24 color blocks. Using this method, we provided matrices for several most commonly used smartphone models. We will keep updating the supported phone models in the option. For functions that are not included in the current APP, we left an entrance for users to conduct their self-calibration of smartphones and illuminations, the calculation follows the same steps above that have been built into the SpeCamX.

The app was developed on Android 11 platform (Google Inc., CA). Currently, we installed it in the Google Pixel 4 smartphone (Google Inc., CA) for demonstration, but any type of smartphone can be used. The development of the APP was conducted in an open-source integrated development environment (IDE): Android Studio (Google, CA). The Weiner estimation algorithm and default transformation matrices developed on Matlab R2021b were incorporated into the SpeCamX. The "Imager" fragment can invoke the built-in camera and use it under default settings (Resolution: 2268X4032; f/1.7; shutter speed: 1/60; white balance: 5500K).

The whole app platform consists of four functional fragments, with their interfaces illustrated in Fig. 2. The "Imager" fragment (Fig. 2a) allows the users to set up required functions before imaging. The users can choose a default setting by selecting the model of their smartphones or enter the customized model to select the type of color chart for recalibration if the smartphone is not in the default list. Fig. 2b shows the drop-down menu of options in the "Imager" fragment. Fig. 2c shows the recalibration page. After the setting, users can launch the imager by tapping the camera icon in Fig. 2a. All the recorded datasets can be searched in the "Records" fragment (Fig. 2d). The "MSI" fragment (Fig. 2e) allows the user to scrutinize the reconstructed spectral images of the subject if needed. Embedded algorithms in SpeCamX automatically process the spectral data cubes to extract the features of interest and present the results in the "Analysis" fragment (Fig. 2f). For example, when the smartphone is used as a bilirubinometer, the "BILI" represents the prediction value made from the region of interest (ROI) in the "Bili" page. Users can acquire these predictions either in the real-time preview mode or after imaging. Except for BBL, other functions like blood perfusion and pigmentation, can also be accessed in the drop-down menu on the top of the interface. The workflow details of these fragments can be found in the Supplementary S1. The spectral imaging quality of SpeCamX has been tested and shown in the Supplementary S2. Overall, the SpeCamX possesses five key features: 1) No external attachments or internal modifications to the smartphone are necessary, thus SpeCamX can be distributed to the users just like other apps in Google shops for example; 2) No further calibration is required when used in the default mode, but there is an option provided for users to calibrate their specific smartphones in a customized mode, which is particularly useful under un-controlled illumination conditions; 3) The customized recalibration can be simply realized by imaging commercialized standard color charts; 4) No offline operation is required, the imaging, processing and analysis functions are all integrated into the SpeCamX; 5) Except for the BBL detection, other functions, like mapping of blood perfusion and pigmentation, can be provided as well. These characteristics would ensure the compatibility and practicability of SpeCamX in mobile health applications.

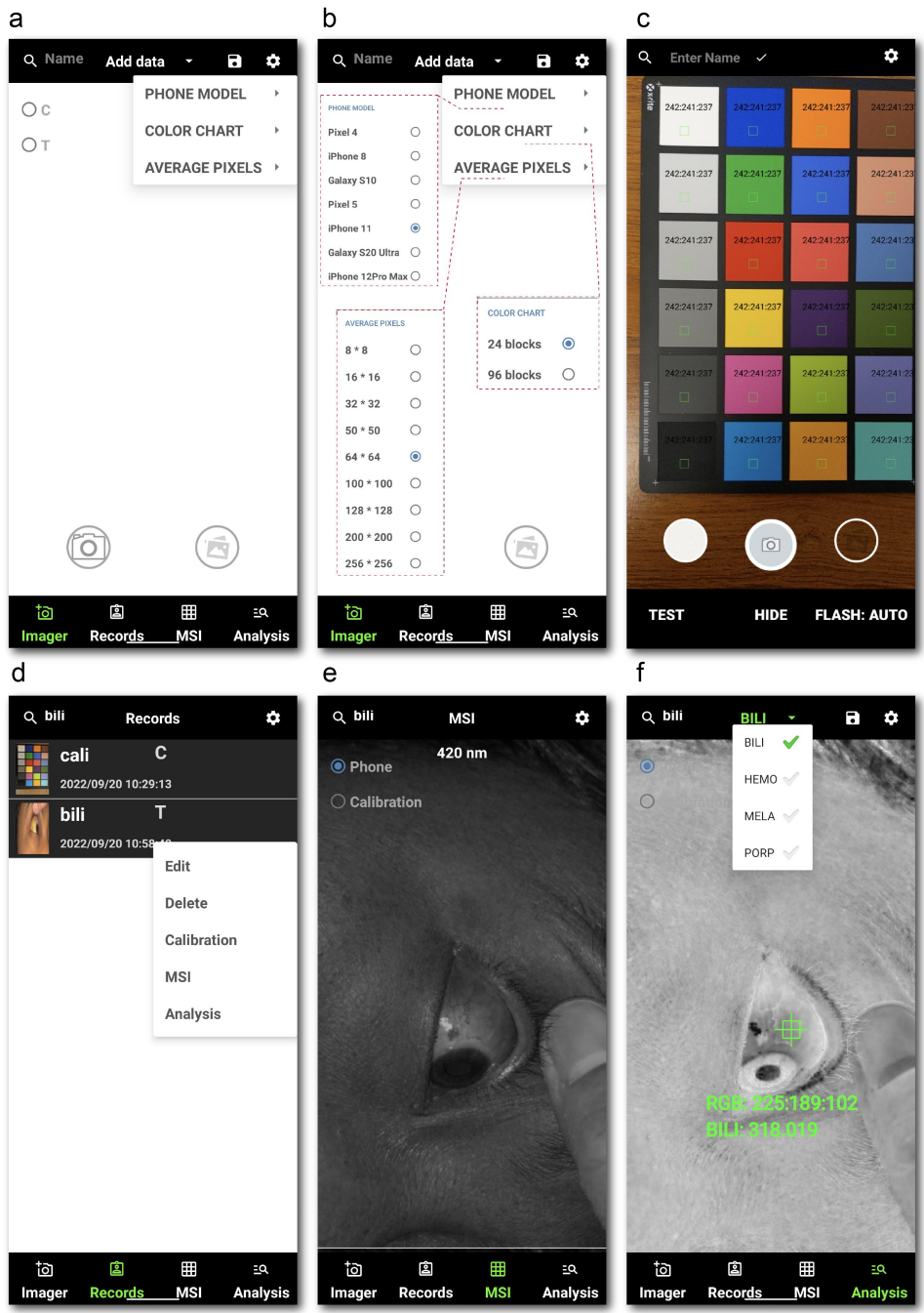

**Fig. 2 Interface and functions of SpeCamX,** which consists of four fragments. **a** "Imager" fragment to set up the camera in the smartphone. **b** Supported functions in the setup options. **c** Recalibration page to generate a customized transformation matrix with a standard color chart. **d** Interface of the "Records" fragment, where users can check the list of acquired data. **e** "MSI" fragment for the presentation of the acquired spectral images of subjects. **f** "Analysis" fragment to present the quantitative information of the extracted feature.

**Machine learning model**

Machine learning is increasingly applicable in medical context because of its excellent ability to recognize subtle pattern features on datasets.[28] The ability of algorithms to extract features which may not be sensitive to the human observer has been extensively explored for a wide variety of predictions including blood tests,[29,30] Alzheimer's disease[31] and cardiovascular diseases,[32] *etc*. Here, the rich but subtle information due to the chromophores embedded within the multispectral images acquired by a smartphone would provide an excellent opportunity to develop a machine learning method to predict the concentration of that chromophore. We therefore developed a machine learning strategy embedded in the SpeCamX for the prediction of blood bilirubin levels. Below, we present the measures of accuracy for the models constructed to create this prediction and contrast the interpretability of different algorithms.

We compared SAL and RGBL in four commonly used machine learning regression algorithms, including ANN, SVM, KNN and RF. We applied these regression algorithms by invoking functions in the *Statistics and Machine Learning Toolbox* in Matlab R2021b. In ANN algorithm, we used *fitrnet* to train a ten-layer, feedforward, fully connected neural network to predict BBL. Function *fitrsvm* fits a support vector regression model. Function *kNNeighborsRegressor* estimates the values of a continuous variable (target) based on one or more independent variables (predictors) based on KNN algorithm. In the RF algorithm, we used *TreeBagger* to combine the results of many decision trees to provide predictions. While designed with different rationale and architectures, these four algorithms were selected because they represent commonly used machine learning methods and are all appropriate for training regression models. In this way, the generalizability of SpeCamX-enabled augmentation can be tested. Afterwards, a hybrid regression model was built by linearly combining these models and incorporated in SpeCamX since hybrid machine learning has the advantages in reducing biases and increasing accuracy.[33,34] The output of the hybrid model can be previewed as the "BILI" value in the "Analysis" fragment of SpeCamX.

**Statistical analysis**

To assess the prediction quality, Pearson's correlation coefficients were calculated between the blood test bilirubin levels and the prediction using SAL and RGBL. Correlations were considered significant if $p<0.001$. 95% prediction bands were computed for the correlation plots. The bias (MD) and 95% limits of agreement (LOA) were computed for the Bland-Altman plots. LOA was computed as 1.96 times the standard deviation of the error. To create ROC curves, 320 cases were classified into positive and negative groups using 17.1 μmol/L as the threshold. All the plots and curves were generated using Origin 2021b.

To obtain a representative reflectance spectrum of the sclera region, first, a spectrum was averaged from a 100*100-pixel$^2$ ROI. Then, a new spectrum was averaged from 10 ROIs in the same snapshot. Finally, from 10 snapshots, another round of average was conducted to obtain the final spectrum. RGB values were measured from the sample ROIs and average calculation. The pupil, large blood vessels and hyper-reflection region need to be excluded from the selected ROIs.

To compare the prediction performance of models using SAL and RGBL, default settings and parameters of functions were used in both methods to keep the consistence of conditions. In RGBL, averaged RGB values from ROIs were used as the input dataset. In SAL, the reflectance spectra averaged from the reconstructed multispectral data cube were used as the input. To avoid overfitting, the prediction of each group was obtained by ten-fold cross-validation. 320 cases were randomly split into ten sets. In each round, the model was trained using nine of ten sets, and then tested on the remaining one set. In ten rounds, every data set needed to be tested for once. The final prediction was averaged from predictions of ten rounds.

To challenge the prediction models with reduced data feeding, the sample subjects were randomly selected by an algorithm developed on Matlab R2021b. We tested sample sizes with data resampling percentage changing from 12.5% (n=40) to 100% (n=320) with a step width at 6.25% (n=20). Under each sample size, RGBL and SAL models were trained with the same dataset. To avoid overfitting, the prediction of each group was obtained by ten-fold cross-validation as well.

**Role of the funding source**

The funder of the study had no role in study design, data collection, data analysis, data interpretation, or writing of the report.

**Results**

Fig. 3 shows the results of SpeCamX-enabled imaging on the sclera in anterior segment of eye (bulbar conjunctiva region[35]) in two representative clinical cases. Clinically, BBLs in the patients were measured at 27.0 and 368.9 μmol/L, respectively. From RGB photographs in Figs. 3a-b, we can clearly observe a darker yellowish color in the sclera of patients with higher BBL. By the side of the photographs, we presented every frame of the spectral images obtained by SpeCamX. In the wavebands from 420 to 480 nm, two cases show distinct signal strength differences in the sclera due to different levels of bilirubin concentration. With a further increase of the wavelength, the difference gradually decreases because the absorbance of bilirubin becomes negligible at the longer wavelengths. In the red bands above 650 nm, no significant absorption can be observed in both cases. Figs. 3c-d illustrate the generation of the reflectance spectra averaged from two imaging trials. In each trial, we acquired ten snapshots at different regions of the sclera. From each snapshot, we calculated an averaged spectrum from the selected ROI. The final reflectance spectra were then averaged from these ten measurements, showing as the black curves. The reflectance spectra also support our above observation that the sclera tissue of the patient with higher BBL shows lower reflectance in wavebands from 420 to 480 nm. Using the captured reflectance spectra, SpeCamX predicted the BBL of these two cases to be 30.5 and 380.5 μmol/L, respectively, which agreed well with the clinical testing results (see the prediction method in the next section).

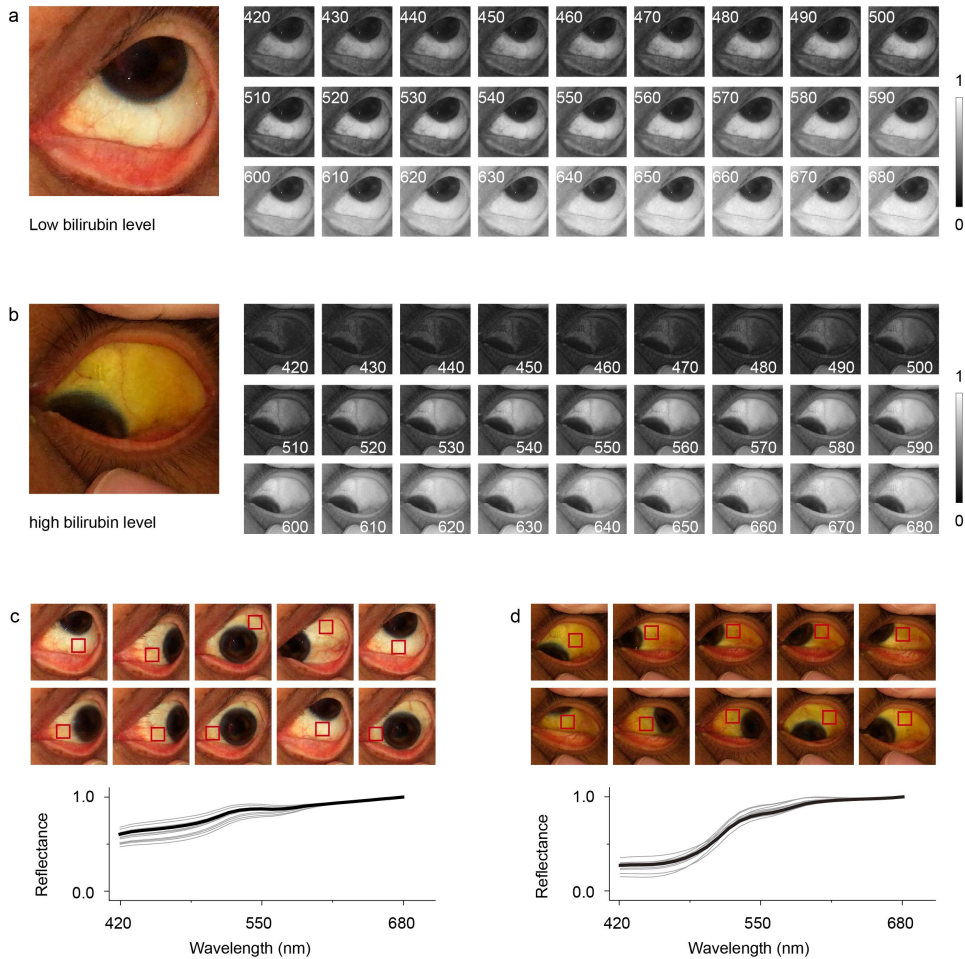

**Fig.3 Examples of clinical imaging using SpeCamX as a bilirubinometer. a, b,** RGB and spectral images of cases with BBLs at 27.0 and 368.9 μmol/L, respectively. **c, d,** 10 samplings on different regions of the sclera and the calculation of the reflectance spectrums of cases in **a** and **b**, respectively. The case with high BBL shows darker yellowish in the RGB photograph and lower reflectance in wavebands from 420 to 480 nm than the case with low BBLs.

Figs. 4a-c show the predictions obtained from the hybrid regression model incorporated in SpeCamX. In 320 cases, the model provided an excellent correlation between SpeCamX prediction and clinical BBL measurements with an R-value above 0.90 ($p<0.0001$). From the Bland-Altman plots (Fig. 4b), we can also observe small limits of agreement (LOA) (+119.90/-117.45 μmol/L) and bias (1.23 μmol/L). Fig. 4c shows the receiver-operating characteristic (ROC) curve of SpeCamX-enabled BBL prediction. The area under the ROC curve (AUROC) was calculated to be 0.97, indicating that SpeCamX and its built-in prediction model can provide a reliable measurement of the BBL by simply taking color photos of the sclera tissue using a smartphone.

To validate our assumption that SpeCamX-enabled multispectral imaging can better predict BBL than conventional smartphone imaging, we compared the quality of prediction using SpeCamX-enabled SAL and RGBL. Besides, given richer information with higher spectral resolution, SAL should also learn quicker than the RGBL model. To validate this point, we also tested and compared the prediction performance while reducing the data feeding. Figs. 5d-e show the predictions and Bland-Altman plots using SAL and RGBL, respectively. In the whole set of 320 cases, RGBL produced visually similar plots with SAL (Fig. 4d), but its Bland-Altman plots indicate a wider LOA (+136.63/-126.57 μmol/L) and bigger bias (5.03 μmol/L) (Fig. 4e). Further, when the sample size was reduced (Fig. 4d), the stability of SAL prediction remains relatively constant, whilst a deterioration of the RGBL prediction is observed (where the regression curve is seen gradually deviating and the prediction band is becoming wider). The corresponding Bland-Altman plots in Fig. 4e validated this observation. With the sample size decreased to 25% (n=80), the LOA of SAL prediction is +114.22/-107.87 μmol/L, but the LOA of RGBL prediction is +166.69/-148.82 μmol/L. We summarized the values of correlation (R), mean difference (MD) and standard deviation (STD) of MD in Fig. 4f-g. Overall, the SAL prediction shows higher R, lower MD and STD than RGBL in all groups.

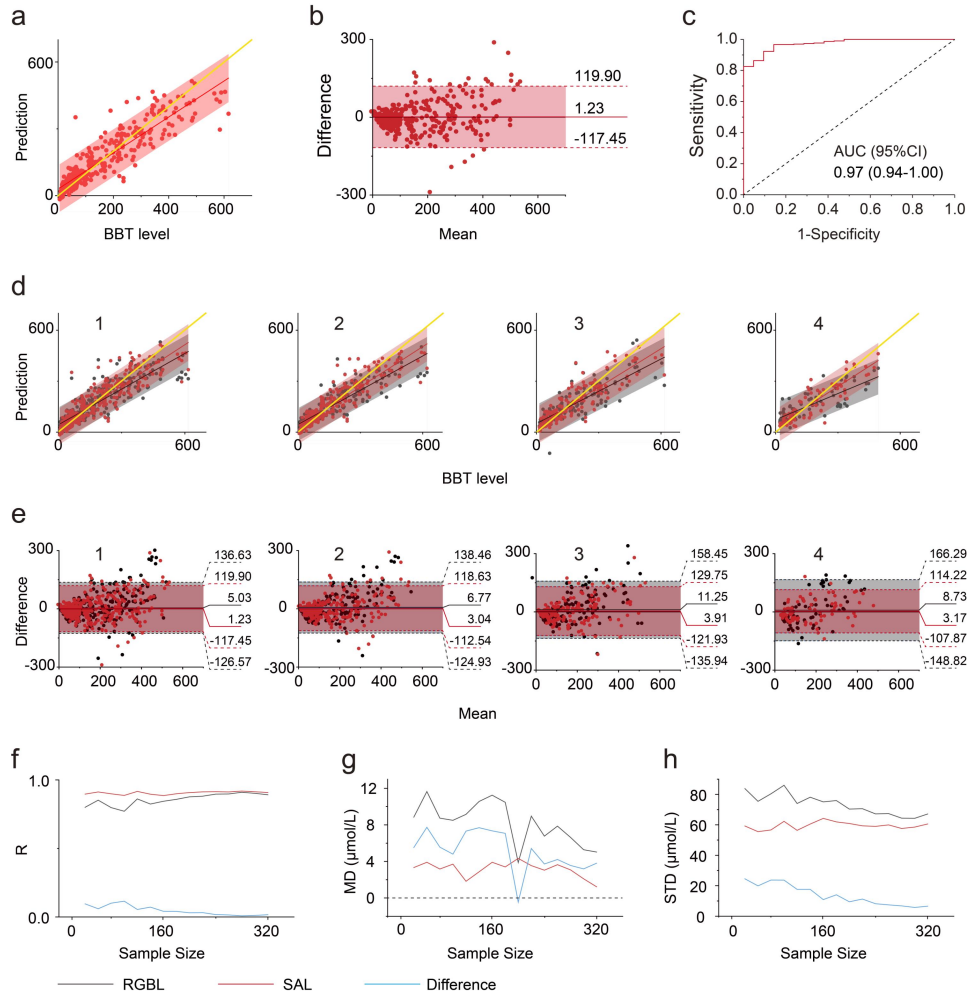

**Fig.4 Quantification and improvements of SpeCamX-enabled SAL prediction using the hybrid regression model. a,** the relationship between BBL and predictions using SAL. Red point: prediction using SAL; red area: 95% prediction band of prediction using SAL. **b,** Bland-Altman plots of predictions using SAL. Red dotted line: 1.96 limits of agreement; red line: MD of prediction. **c**, the ROC analysis of prediction using SAL. A bigger AUROC indicates a better diagnosis. **d,** the comparison between predictions using SAL and RGBL in different sample sizes (resampling percentages from left to right: 100%; 75%; 50%; 25%). Here RGBL was conducted using the same hybrid model as SpeCamX. Red point: prediction using SAL; red area: 95% prediction band of prediction using SAL. Black point: RGBL prediction; black area: 95% prediction band of RGBL prediction; **e,** corresponding Bland-Altman plots of predictions in **d**. Black dotted line: 1.96 limits of agreement of RGBL prediction; red dotted line: 1.96 limits of agreement of prediction using SAL; black line: MD of RGBL prediction; red line: MD of predictions using SAL. **f-h,** Quantification of R (**f**), MD (**g**) and STD of MD (**h**) between BBLs and prediction using SAL/RGBL along different resampling percentages from 12.5% to 100%. Black curve: RGBL prediction; red curve: predictions using SAL; blue curve: the difference between RGBL prediction and predictions using SAL.

The enhancement of SAL prediction over RGBL is benefited from the multispectral information brought by the SpeCamX, rather than by a specific design of the prediction model. To demonstrate this point, we further separately investigated the quality of predictions using SAL and RGBL implemented by individual

ANN, SVM, KNN and RF algorithms. The input of SAL and RGBL is the spectra saved in SpeCamX and corresponding RGB values of ROIs. The predictions were detailed in Supplementary S3 and S4. To summarize these comparisons, SAL improved the prediction quality to varying degrees, especially with less data feeding. To illustrate this point clearer, we quantified the prediction performance of SAL and RGBL with data resampling percentage ranging from 12.5% (n=40) to 100% (n=320) with a step width at 6.25% (n=20). The R, MD and STD were then measured and presented as curves in Fig. 5. The evolution curves showed that the R of SAL always remained at high levels of 0.9 ($p<0.0001$), even when only 12.5% (n=40) of the data was used to train the model. On the contrary, the R of RGBL can be even lower than 0.6. In all methods except for SVM, the prediction biases of SAL are close to 0, smaller or at least comparable to RGBL predictions. In SVM, the bias of SAL is relatively increased, but still averagely 55% (16.87±3.27 μmol/L) smaller than that of RGBL. The STD of MD in SAL slightly increases with smaller sample size but overall lower than 75 μmol/L, which is almost the best level the RGBL can achieve. Figs. 5e-f show that the AUROC value of SAL in all models are above 0.94, which outperforms that of RGBL in all algorithms. Besides the enhancements exist in absolute values, the red SAL curves in Fig. 5 seem to perform higher stability in all groups than RGBL curves. We quantified the standard deviations of these four indices in all groups with different sample sizes and algorithms (Supplementary S5). The standard deviations of R, MD and STD of MD in SAL prediction are 72% (0.06), 50% (6.80 μmol/L) and 66% (11.74 μmol/L) higher than those in RGBL prediction. All these merits demonstrated that SpeCamX-enabled SAL can significantly augment the BBL prediction quality regardless of the learning algorithms used, which indicates that its superiority is originated from the multispectral imaging capacity in SpeCamX. In this case, it is reasonable to expect that any machine learning augmentation can be added when deploying SpeCamX for applications other than bilirubin level detection as demonstrated here.

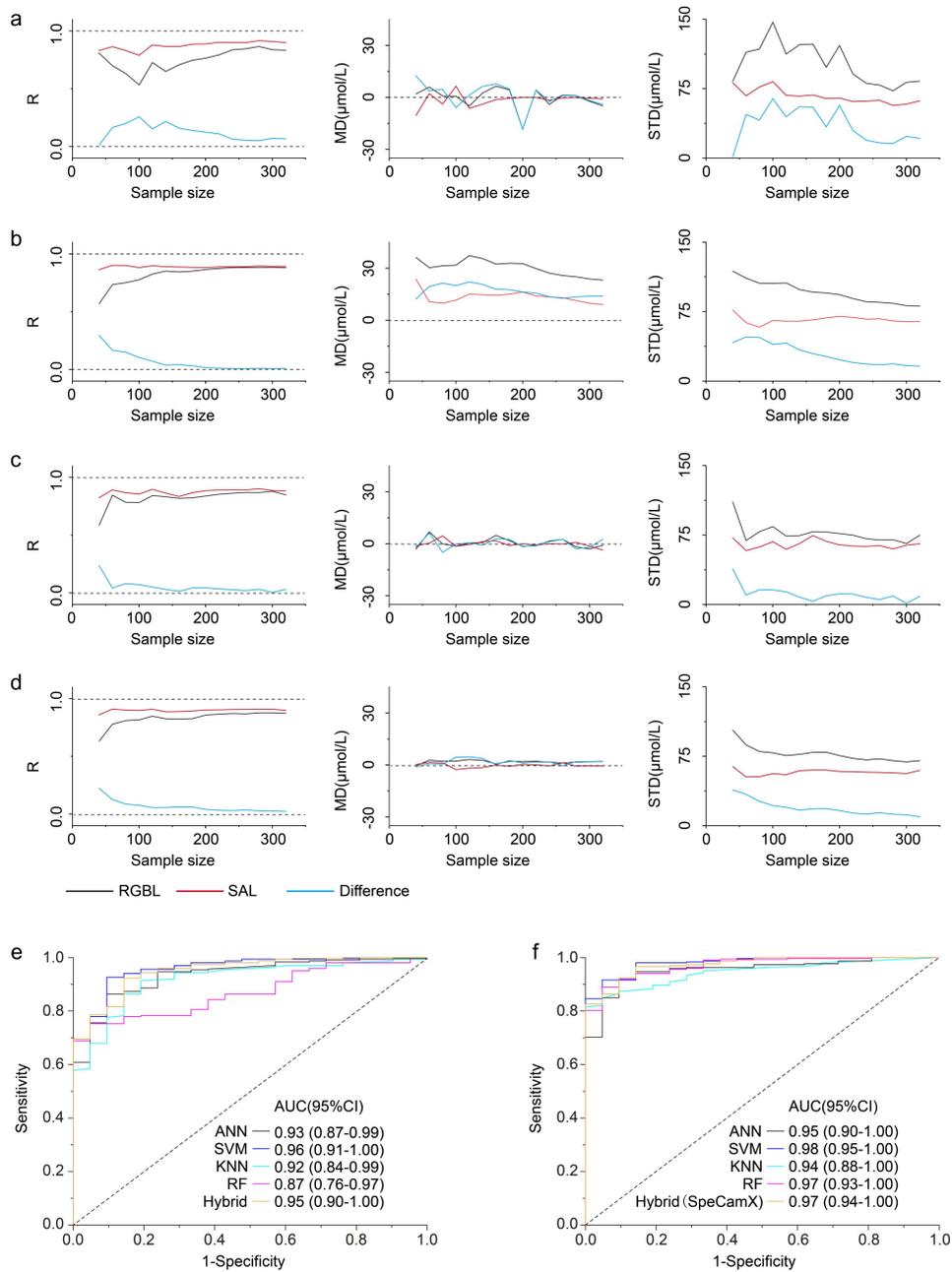

**Fig.5 Quantification and improvements of SpeCamX-enabled SAL prediction using single machine learning models**. **a-d,** Quantification of R (left), MD (middle), and STD (right) of the BBLs predicted with SAL/RGBL when data feeding varied from 12.5% (n=40) to 100% (n=320) into different machine learning models: **a**, ANN; **b**, SVM; **c**, KNN; **d**, RF. Black curve: RGBL prediction; red curve: SAL prediction; blue curve: difference between RGBL prediction and SAL prediction. **e-f,** the receiver-operating characteristic (ROC) analysis of prediction using RGBL (**e**) and SAL (**f**) using the single and hybrid machine learning model. Regardless of single or hybrid models, SAL achieves a better AUROC than RGBL, which indicates a better diagnostic value for SAL-based SpeCamX. All the groups with different sample sizes are randomly selected from the data pool of 320 patients.

**Discussion**

The development of mobile health based on smartphones facilitates daily monitoring of many vital signs and body tissue compositions, some of them have the potential to indicate disease conditions that are difficult to detect/diagnose without physically visiting healthcare providers.[36] Exploring and developing such techniques can clearly promote and benefit public healthcare. For example, the global incidence of liver disease (LD) is estimated at 1.5 billion, which leads to about two million deaths each year.[37,38] Close monitoring of at-risk populations is believed to be an effective strategy to control its progression and spread.[39] However, frequent testing through visiting the clinical labs imposes an inevitable burden on patients, both psychologically and economically, impacting the compliance to seek for medical services. To improve the clinical compliance and promote their willingness to accept the monitoring of liver health conditions, one solution is to noninvasively detect bilirubin levels in the serum, preferably that can be performed in a non-clinical environment. The balance of BBL in circulation relies on a normal liver metabolism, which makes it a suitable biomarker of liver functions.[40,41] At different severity stages of LD, bilirubin dysbolism accumulates and eventually causes different levels of hyperbilirubinemia, which usually appears as the yellowish pigmentation in body tissue.[42,43] With distinct optical spectral properties, bilirubin-induced pigmentation is suitable to be noninvasively measured using optical sensors to estimate the BBL and finally predict the liver condition. Some of these sensors, like transcutaneous bilirubinometer, equip spectral illumination for the detection of the light absorption to estimate BBL.[44] We also see some portable versions of these sensors that are developed to save people from the frequent clinical visits,[45] however, substantial investment is still required to acquire these dedicated devices, which eventually prevents them from wider spread use.

Herein, we provided a strategy by developing a mobile application termed SpeCamX to transform smartphones into spectral imagers and tested this strategy in BBL prediction. SpeCamX can acquire multispectral data cubes through a single snapshot using a smartphone without external attachments. Besides, SpeCamX provides a set of pre-stored transformation matrices for some popular smartphones and standard room conditions. This provision would save users from the requirements on the expertise of complicated color calibration and off-line processing. Moreover, after spectral imaging, SpeCamX uses the "Analysis" fragment in the app to analyze the chromophore information from acquired multispectral data cubes. Except for the bilirubin prediction discussed in this study, the SpeCamX also provides options to map the blood perfusion, melanin pigmentation and other chromophores within human tissue. Given that smartphones are already integrated terminals equipped with detectors, processors and display units, the installation of SpeCamX will enable them to conduct selfie health screening and monitoring with high independence and almost cost to nothing.

In this study, as a proof of concept, the output data cube of SpeCamX was set to be in 27 channels covering from 420 to 680 nm with a step width at ten nanometers. We assumed that bilirubin and hemoglobin dominated the color of scleral tissue in the bulbar conjunctiva region (which is easily accessible by smartphone), thus the selected wavelength bands covered both the absorption peaks and troughs of these two chromophores. Moreover, the format of output data, including the spectral resolution and channel number, can be adjusted according to specific sample compositions and processing methods. For instance, the user can set up a more refined spectral resolution (i.e. to have denser spectral channels) around bilirubin's absorption peak to maximally minimize the effect from hemoglobin content. Another example is in developing machine learning algorithms, a setting with more channels and higher spectral resolutions would be particularly useful to enable potential features to be fully learned. Furthermore, SpeCamX can also support non-learning algorithms by imaging at typical wavelengths, like 460 and 500 nm used in transcutaneous bilirubinometer. Overall, SpeCamX-enabled medical imaging possesses a high technical flexibility for further optimization and exploitation.

Through the clinical imaging of patients with LD, we demonstrated the enhancements in predictions realized by SpeCamX. Encouraged by this study, we believe SpeCamX may also act as an effective tool to monitor other bilirubin-related diseases, like neonatal jaundice. In fact, both infants and guardians should benefit from the real-time, non-contact and non-invasive evaluation mode of this SpeCamX-augmented smartphone bilirubinometer. Further, if we see the bilirubin prediction described here as evidence, it should be reasonable to expect similar prediction quality when evaluating other chromophores. For example, SpeCamX should conceivably better describe hemoglobin behaviors than regular camera apps, which would strengthen the monitoring of blood and vascular abnormalities. In a word, SpeCamX has the potential to contribute to conventional smartphone-enabled medical imaging methods and improve their performance.

There are still some limitations in our study. Although we maintained the diversity of patients in gender and age in the demonstration of SpeCamX augmented bilirubinometer, only Asian patients were enrolled and imaged. Being dominated by bilirubin and hemoglobin, the color of scleral tissue is not correlated with the complexions in subjects, but further investigation is needed. Besides this, the property of SpeCamX to estimate targets from more complicated backgrounds remains untested. Some human tissues may contain chromophores with more diversity and less predictability. Aiming to spread the usage of SpeCamX, we need to further explore its usability in these scenarios.

Looking forward, though our current study was conducted in clinics, a wider application of SpeCamX is expected for the public in their daily lives. In the future, we will keep working on the optimization, packaging, and approval of this app and the method to provide its open access. Through this way, we may conduct studies on the health care of a larger population with more diversity in regions, races, and complexions. Meanwhile, we will keep updating the embedded "Analysis" fragment by uploading more functions to empower this platform for the screening and monitoring of more health issues.

## Acknowledgements

This work was supported by Washington Research Foundation. The funding organization had no role in the design or conduct of this research.

## Contributors

QH was responsible for the conceptualization, formal analysis, methodology, project administration, software, validation, visualization and writing-original draft. WL was responsible for the data curation, formal analysis, resources, project administration and supervision. YS was responsible for the software. YY was responsible for the supervision. YZ was responsible for the formal analysis. WG was responsible for the data curation. ZS was responsible for the conceptualization, data curation, formal analysis, resources, methodology and project administration.

## Declaration of Interests

All authors declare no competing interests.

## Data sharing

The data that support the findings of this study are available from the corresponding author, but restrictions apply to the availability of these data, which were used under license for the current study, and so are not publicly available. Data is however available from the corresponding author upon reasonable request. The code and software that support this work is copyright of the Regents of the University of Washington and can be made available through license.

# Supplementary information:

# Augmented smartphone bilirubinometer enabled by a mobile app that turns smartphone into multispectral imager

Qinghua He[1, 2, †]; Wanyu Li[3, †]; Yaping Shi[2]; Yi Yu[1]; Yi Zhang[2]; Wenqian Geng[3]; Zhiyuan Sun[1, *]; Ruikang K Wang [2, 4, *]

[1]Changchun Institute of Optics, Fine Mechanics and Physics, Chinese Academy of Science, Changchun, Jilin 130033, China

[2]Department of Bioengineering, University of Washington, Seattle, Washington 98105, USA

[3]Department of Hepatobiliary and pancreatic Medicine, The first Hospital of Jilin University NO.71 Xinmin Street, Changchun, Jilin 130021, China

[4]Department of Ophthalmology, University of Washington, Seattle, Washington 98109, USA

[†] These authors contributed equally to this work.

[*] sunzhiyuan@ciomp.ac.cn

## S1. Workflow of SpeCamX

*"Imager" fragments*

The subject is labeled from the left top of the interface, which is auto-saved together with date information in the filename of acquired data cubes for further references. According to different illumination conditions, we provided several options to set up the camera (see main text Fig. 2b). The default condition is to use built-in flashlight to illuminate subject. In this case, users can select their smartphone models in the "PHONE MODEL" option under the drop-down menu of settings. In this option, we pre-stored transformation matrices (TMs) to support a variety of smartphone models. Therefore, users can invoke corresponding matrices by simply selecting the smartphone model they use. As long as it is in dark environment, you are ready to go. In our clinical imaging test, the data were acquired using this setup.

In addition to the default setting, there are a number of other options for users to select if the requirement for default setting is not met, for example the smartphone that is in use is not supported or the environment is not in dark condition. Under these circumstances, the user can generate a customized TM on site through a self-recalibration step. In doing so, the user needs to first tap the "COLOR CHART" icon, in which there are provided with two options, i.e. either to use "24 blocks" or "96 blocks" to build the TM matrix (see main text Fig. 2b). Here, we used X-rite ColorChecker as the default color standard example for recalibration. The "24 blocks" option is provided for color charts with 24 classic color series. The available products include *X-rite ColorChecker Classic/Passport/Mini/Nano* (**S-Fig.1**). The "96 blocks" option works for *X-rite ColorChecker Digital SG* which includes expanded 96 color blocks with standard reflectance spectrums (**S-Fig. 2**). After selecting the color chart option, the user can tap into a calibration page and sample the on-site color chart by tapping the camera icon. In this page, a box array would be generated to guide the sampling of color blocks. For example, in the page of "24 blocks", a 6 by 4 green box array would appear on the interface (see main text Fig. 2c). By pressing the photo button, the averaged RGB values in these boxes would be collected to compute a new TM with their reflectance spectra. During the calibration, the RGB values can be previewed on the top of corresponding boxes to prevent overexposure. The procedures for "96 blocks" option are the same but the box array is set to be 12 by 8.

Furthermore, even under an illumination that cannot be controlled or stabilized, there is provided another option to perform real-time recalibration in the "COLOR CHART" tap, which is termed "co-illumination" strategy. In the "co-illumination", the color charts and subjects will have to be placed within the same field of view for imaging, which will guarantee a shared illumination condition between the calibration and testing data cubes (**S-Fig. 3**). For this purpose, we can manually adjust the relative positions and sizes of the sampling box array to avoid the overlap between the color chart and the subject in the field of view. This strategy is suitable for trials if it is not possible to control the illumination environment, for example in the outdoor environment.

Except for on-site imaging, users can also load previous or external data by tapping the "Add data" icon on the top of the interface. The acquired and uploaded data can be labeled with "C" and "T", representing data for color charts and subjects, respectively. In this case, every smartphone installed with SpeCamX can act as a processing platform for an external data set.

*"Records" fragments*

The data specific to a subject can be recalled and reviewed by selecting that subject or sliding the drop-down menu up and down to check the thumbnails. To edit the data, several operations, including "Delete", "Calibration", "MSI" and "Analysis" are offered by long clicking. "Calibration" is provided for color chart data labeled with "C" to calculate a new TM. Tapping "MSI" and "Analysis" can switch to the corresponding fragments to show the related information of the selected case.

*"MSI" fragments*

The grayscale images at wavelengths from 420 to 680 nm are presented in a scrolling display started with the corresponding RGB photograph (**S-Fig. 4**). The wavelength is labeled on the top of each spectral image concurrently.

*"Analysis" fragments*

The BBL estimation results would be mapped in the "Bili" page, where a green box would guide the users to select a ROI. The averaged reflectance spectrum from the ROI would be read out that provides the prediction of BBL. Considering the complexity of clinical imaging, we provided enough flexibility for the sampling procedure. The size of the ROI can be set in the "AVERAGE PIXELS" option (**S-Fig. 5**) and its position can be manually shifted to the desired regions of interest. (**S-Fig. 6**). Averaged RGB values in the ROI is shown in display so that the user can check to make sure the region is not overexposed. The shown "BILI" value represents the prediction made by the reflectance spectrum averaged from the ROI. The spectrum and predicted BBL are saved by tapping the photo icon for later use.

Except for the prediction of BBL (Bili) used in this study, SpeCamX also integrated algorithms to map other features like blood perfusion (Hemo), melanin pigmentation (Mela) and et al. Users can launch these functions by switching analysis algorithms from the drop-down menu on the top of the interface.

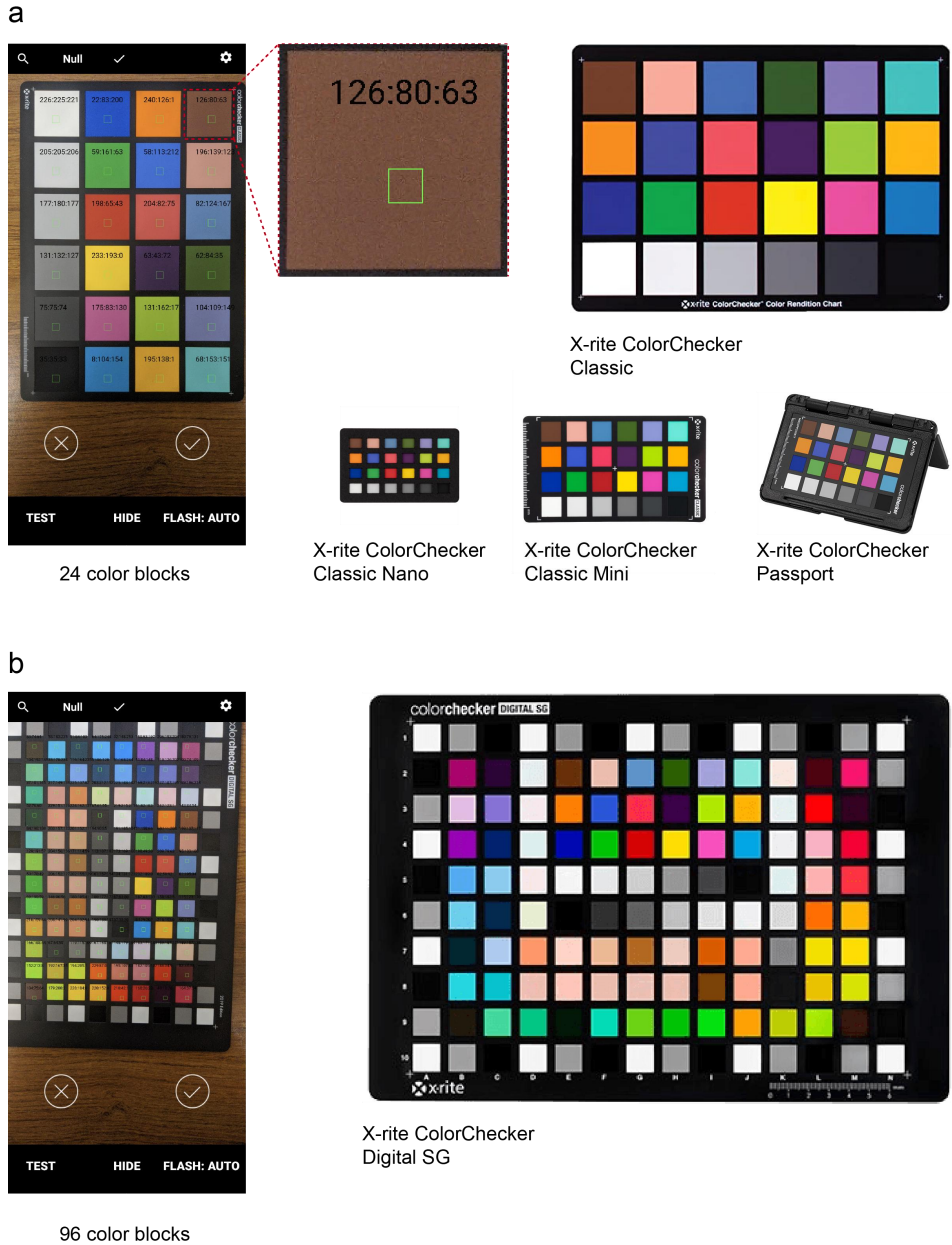

**S-Fig. 1.** Interfaces of the "24 blocks" and "96 blocks" calibration pages and compatible color charts for these options. **a**. "24 blocks" option for 24 classic colors. A 6 by 4 box array was previewed to guide the sampling of RGB values. The RGB values were previewed on the top of the sampling box. Standard color charts with 24 classic colors, like *X-rite ColorChecker Classic/ Passport/ Classic Mini/ Classic Nano*, can be used in this option. **b**. "96 blocks" option for expanded colors. A 12 by 8 box array was previewed to guide the sampling of RGB values. *X-rite ColorChecker Digital SG* can be used in this option.

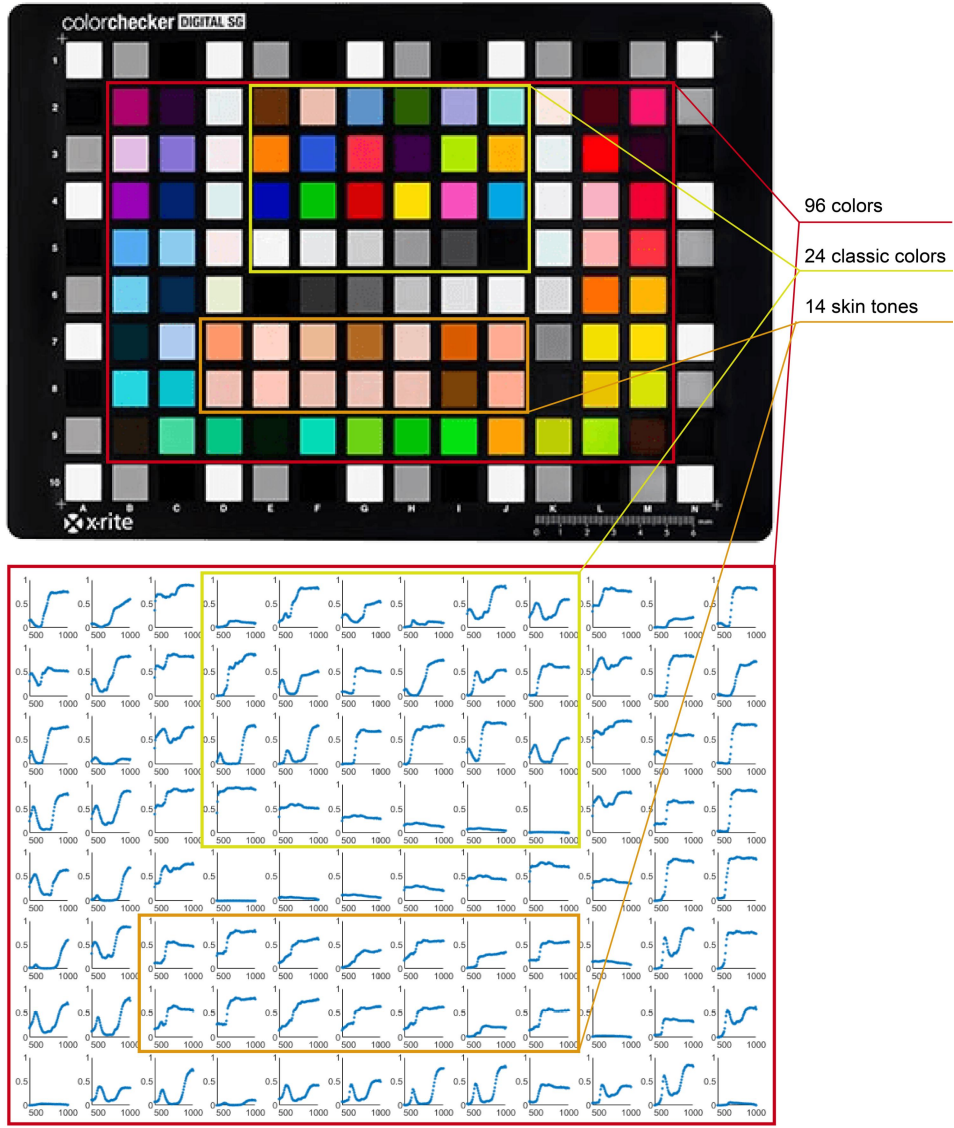

**S-Fig. 2.** Reflectance spectrums of 96 color blocks in *X-rite ColorChecker Digital SG*.

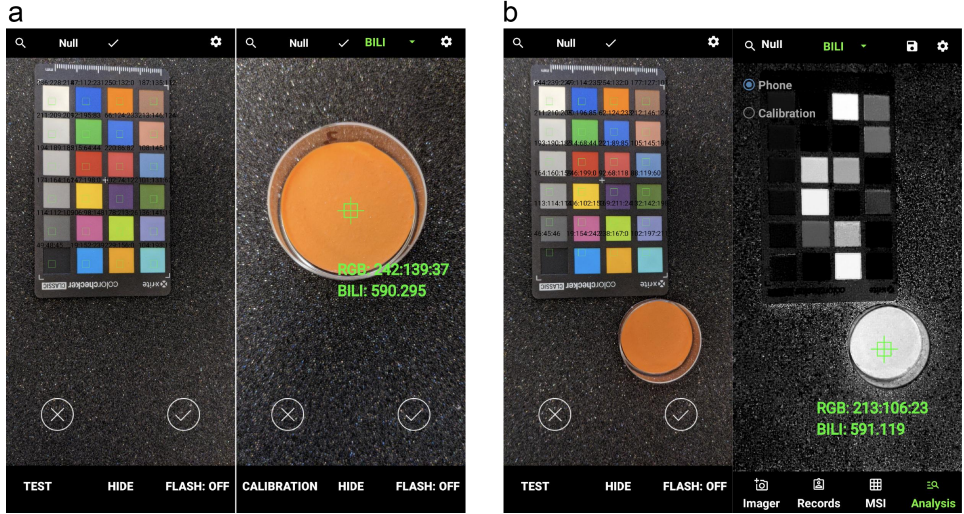

**S-Fig. 3.** Recalibration under uncontrolled and instable illumination. While the external illumination source cannot be controlled and stabilized, users can adopt a "co-illumination" step to realize real-time calibration. Instead of separately conducting "Calibration" and "Test" (**a**), the subject would be imaged with the color chart together in "Calibration". To do this, the box array would be rescaled to make space for the subject (**b**). The prediction results would be presented in the "Analysis" fragment. .

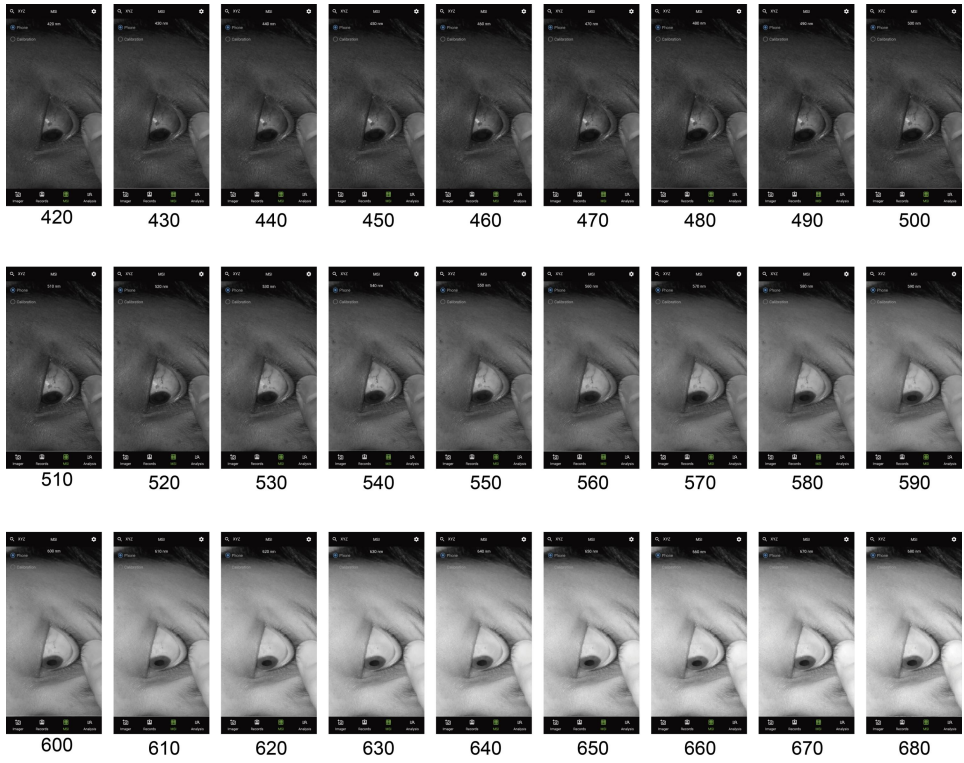

**S-Fig. 4.** The grayscale spectral images at wavelengths from 420 to 680 nm would be presented in the "MSI" fragment. The images can be checked by scrolling the screen, the wavelength would be accordingly labeled on the image.

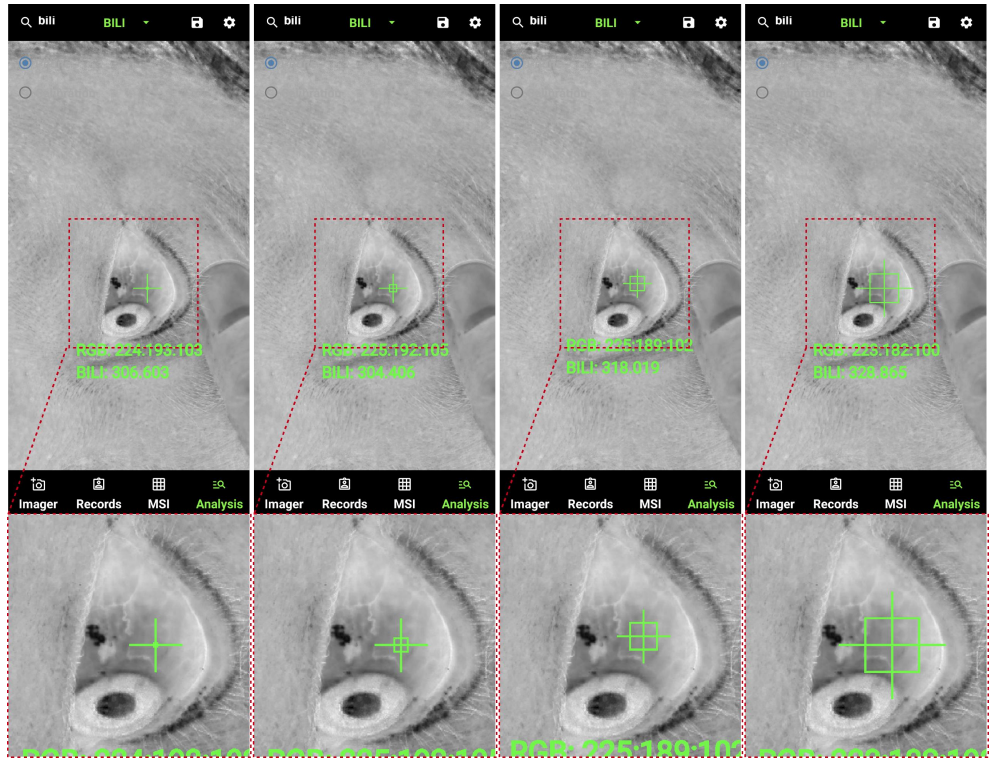

**S-Fig. 5.** The size of ROI can be set up in the "AVERAGE PIXELS". Interfaces from the left to the right were under the setting of "8*8", "32*32", "64*64" and "128*128", respectively.

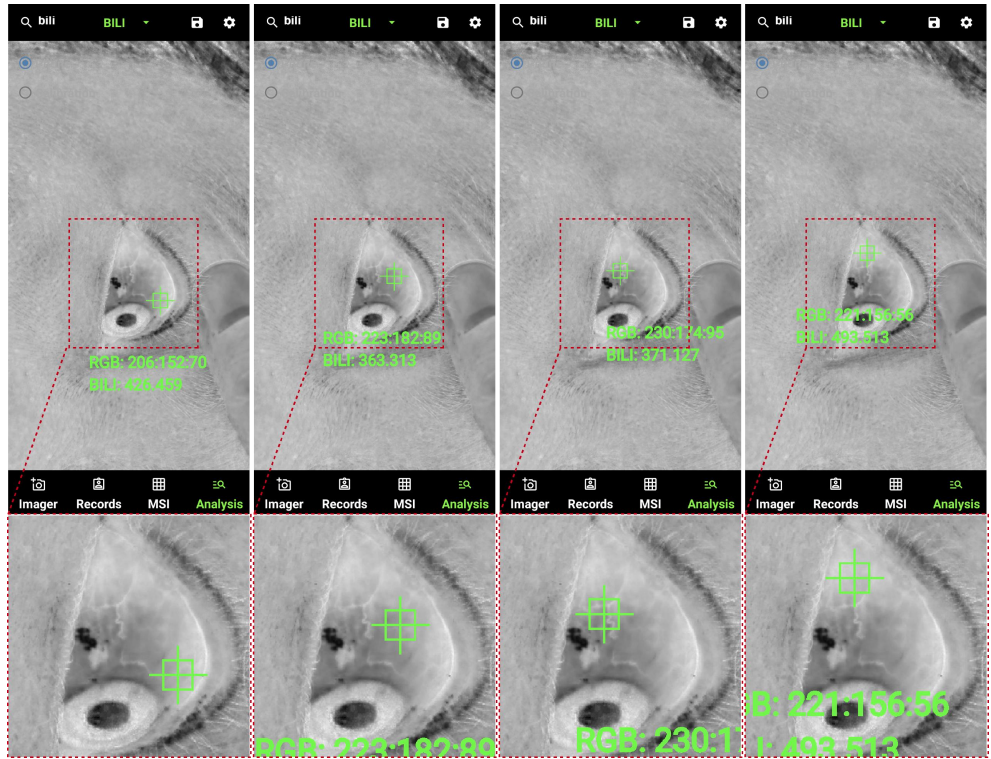

**S-Fig. 6.** Manual adjustments of the ROI position to sample different tissue regions by dragging the green marker.

## S2. Accuracy of spectral reconstruction using SpeCamX

The retrieval of spectral information from an RGB camera relies on the TMs that are obtained during calibration process and pre-stored in SpeCamX. Here we describe the performance evaluation for the spectral reconstruction realized by the pre-stored matrices. In the method, the default TM for the smartphone (Google Pixel 4, Google Inc., CA) was obtained from 24 Macbeth classic color blocks in the calibration step. We used a *X-rite ColorChecker Digital SG* (note that this was not the one used for calibration) as the target for evaluation. After the smartphone took the color photo of the ColorChecker, the app automatically reconstructed the spectra for both classic colors and skin tones. The comparisons of the standard and reconstructed spectra were shown in **S-Fig. 7(a)**. The reconstructed spectra highlighted in red match very well with the references in both classic and skin tone colors. The RMSE of each color block was measured to quantify the reconstruction accuracy (**S-Table 1**). The average RMSE of all color blocks is less than 0.04, indicating an accurate reconstruction of reflectance spectra from 420 nm to 680 nm from a single snapshot RGB image.

RGB values can be influenced by different illumination conditions and channel sensitivities, which may lead to inconsistent responses under different camera settings. To address this issue, SpeCamX provided both default TMs and recalibration options to stabilize the quality of spectral imaging. Here, we simulated some extreme conditions by adjusting the color temperature and ISO of the camera to challenge this stability. The *X-rite ColorChecker Digital SG* was again used in this evaluation and imaged under different settings in camera. The color temperature was increased from 2500K to 9000K with a step width of 500k.The ISO was set to be 880, 840, 800, 720, 640, 570, 500, 450, 400, 360, 320, 285, 250 and 200, respectively. The RGB values and reflectance spectra of all color blocks in these procedures were recorded by SpeCamX. As an example, the results of color block #7D were shown in **S-Figs. 7(b, c)**. The horizontal axis includes both RGB and multispectral imaging (MSI) channels, while the vertical axis represents the settings in camera. The RGB signals were normalized into the same scale as the MSI channels and mapped in the same figure. With the color temperature increased from 2500K to 9000K, we observed the signals in G and B channels remained relatively stable, but that in R channel increased proportionally from the RBG images. In contrast, the SpeCamX provided relatively stable signals in all reconstructed reflectance spectral channels despite the change in the color temperature. A similar result can be observed in **S-Fig. 7(c)**, where RGB values are increased with the increase of ISO but the SpeCamX provided consistent spectral reconstruction in all the channels. To quantify the consistency, we calculated the standard deviations of signals in each channel for all color blocks, shown in **S-Figs. 7(d, e)**, respectively. After normalizing RGB values into the same scale as MSI signals, the averaged standard deviations in RGB channels were calculated to be $0.045 \pm 0.038$ and $0.070 \pm 0.033$ when changing the color temperature and ISO, respectively. The corresponding standard deviation values in MSI channels were calculated to be $0.016 \pm 0.005$ and $0.013 \pm 0.006$. Compared with the RGB values, the signals in MSI channels perform much lower standard deviations. These experiments demonstrate that the SpeCamX can reconstruct accurate spectral information of the sample with a high consistency under different device conditions and settings.

Next, we prepared a set of phantoms to further test the performance of SpeCamX to recover the spectral information of bilirubin. The bilirubin levels were 0.00, 0.23, 0.47, 0.94, 1.88, 3.75, 7.50, 15.00, 30.00 mg/dL in the phantoms one to nine. Firstly, we weighed 10 g agar powder and added them into 100 mL deionized water. The mixture was maintained in a water bath at 100 °C under continuous mechanical stirring. Then, 0.5 g titanium dioxide powders were added into the solution to simulate the optical property of sclera background. After stirring for five minutes, we weighed and added different amounts of bilirubin powders into the mixture to prepare phantoms with different bilirubin concentrations (Phantom 1-9: 0.00, 0.23, 0.47, 0.94, 1.88, 3.75, 7.50, 15.00, 30.00 mg/dL). After stirring for another 10 minutes, the mixture was cooled to 47 °C under continuous stirring and then dumped into a petri dish for cooling and forming. The acquired RGB photographs of these phantoms are shown in the inset of **S-Fig. 7(f)**, appearing deeper yellowish pigments with the increase of the bilirubin concentration. Their reflectance spectra were obtained by the SpeCamX and presented as the curves in **S-Fig. 7(f)**. These spectra were normalized by the reflectance at 680 nm because the absorbance of bilirubin at this wavelength band is negligible. Compared with phantom 1 without bilirubin, other phantoms give lower reflectance around 460 nm and the rate of reduction is similar to that of the concentration. We calculated the values of rate reduction at 460 nm and mapped the points with their bilirubin concentrations, accordingly, shown in **S-Fig. 7(g)**. There is an excellent linear relationship between these two variables, verifying that SpeCamX can be used to detect and quantify the optical absorption of bilirubin.

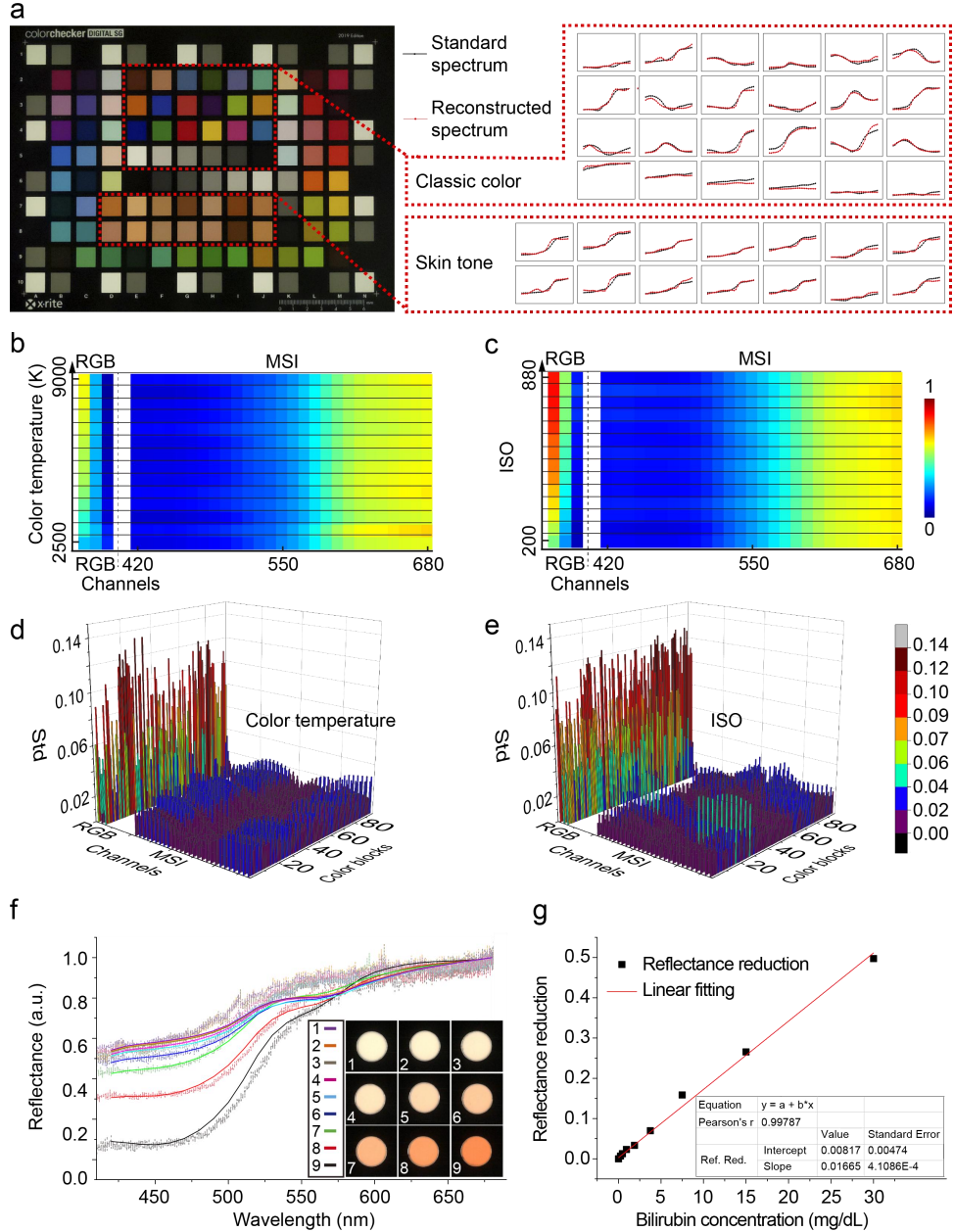

**S-Fig. 7. Performance of spectral imaging using SpeCamX. a** Spectral reconstruction test on a standard color checker (X-rite ColorChecker Digital SG). The pre-stored TM was applied on 24 classic colors and also 14 skin tone colors to estimate the reflectance. The reconstructed spectra were compared with the standard spectra on the right. **b, c** Spectral reconstruction of color block #7D under different ISO and color temperature settings. The color temperature is increased from 2500K to 9000K with a step width of 500k. The ISO is set to be 880, 840, 800, 720, 640, 570, 500, 450, 400, 360, 320, 285, 250 and 200. **d, e** Standard deviations of signals in each channel of 96 color blocks with the alternation of color temperature and ISO settings. RGB values of color block #7D show much higher standard deviations than the reconstructed spectra in this procedure. **f** Spectral reconstruction test on bilirubin phantoms. The concentrations of bilirubin are 0.00, 0.23, 0.47, 0.94, 1.88, 3.75, 7.50, 15.00, 30.00 mg/dL in phantom 1-9, respectively. The inserted figure shows the RGB photographs of phantoms which show progressively darker yellowish pigmentation from phantoms 1 to 9. The curves present the reconstructed reflectance spectra accordingly. The dotted line curves present the reflectance spectra measured by a standalone spectrometer. The reflectance at wavelengths around 460 nm shows gradual decrease from phantom 1 to 9 because of the increased absorption of bilirubin. **g** The linear relationship between the reflectance reduction measured by SpeCamX and the bilirubin concentrations of phantoms. The inserted table shows the fitted equation and fitting errors.

**S-Table 1 RMSE of reconstructed reflectance spectrums using SpeCamX in X-rite ColorChecker Digital SG**

| | B | C | D | E | F | G | H | I | J | K | L | M |
|---|---|---|---|---|---|---|---|---|---|---|---|---|
| 2 | 0.033 | 0.040 | 0.041 | 0.034 | 0.048 | 0.028 | 0.046 | 0.021 | 0.059 | 0.049 | 0.026 | 0.083 |
| 3 | 0.043 | 0.039 | 0.025 | 0.039 | 0.043 | 0.046 | 0.035 | 0.048 | 0.025 | 0.047 | 0.048 | 0.046 |
| 4 | 0.042 | 0.013 | 0.045 | 0.044 | 0.086 | 0.034 | 0.035 | 0.078 | 0.025 | 0.019 | 0.034 | 0.053 |
| 5 | 0.035 | 0.058 | 0.047 | 0.030 | 0.025 | 0.033 | 0.028 | 0.013 | 0.026 | 0.042 | 0.036 | 0.049 |
| 6 | 0.022 | 0.044 | 0.054 | 0.017 | 0.041 | 0.016 | 0.029 | 0.045 | 0.018 | 0.031 | 0.039 | 0.042 |
| 7 | 0.020 | 0.025 | 0.045 | 0.040 | 0.018 | 0.029 | 0.025 | 0.055 | 0.049 | 0.025 | 0.074 | 0.077 |
| 8 | 0.051 | 0.033 | 0.041 | 0.045 | 0.045 | 0.025 | 0.024 | 0.041 | 0.042 | 0.028 | 0.047 | 0.015 |
| 9 | 0.016 | 0.029 | 0.019 | 0.013 | 0.035 | 0.045 | 0.063 | 0.047 | 0.051 | 0.044 | 0.016 | 0.053 |

## S3. Comparison of prediction performance between SAL and RGBL using ANN and SVM

**S-Fig. 8(a)** shows the predictions obtained by ANN-based regression models. In 320 cases, we realized a correlation between SpeCamX and clinical BBL with a R at 0.90 (p<0.0001) and 0.83 (p<0.0001) for SAL and RGBL models, respectively. Except for higher R, the SAL prediction band (red band) is also narrower than the RGBL prediction band (black band). From the Bland-Altman plots (**S-Fig. 8(b)**), we can also observe smaller prediction bias in SAL prediction. Given richer information provided with higher spectral resolution, the SAL model should also learn quicker than the RGBL model. To validate this point, we reduced the data feeding to test and compare the prediction performance as well. **S-Fig. 8(a)** show the prediction results while the subjects were reduced by randomly resampling 75% (n=240), 50% (n=160), 25% (n=80) of the whole 320 cases, respectively. The prediction bands of both SAL and RGBL widened when the data set size was decreased. However, the SAL prediction band kept its relative stability while the RGBL prediction band showed increased instability. This observation was quantifiably verified in their corresponding Bland-Altman plots (**S-Fig. 8(b)**). The limits of agreement (LOA) of SAL prediction changed from +122.71/-124.43 μmol/L to +149.66/-157.26 μmol/L when the case number decreases from 320 to 80, but the LOA of RGBL expanded from +161.28/-171.10 μmol/L to +236.29/-234.65 μmol/L. This difference indicates that SpeCamX-enabled SAL can achieve predictions with higher quality than RGBL when the data feeding is limited.

In the results of SVM-based regression models (**S-Fig. 8(c)**), the regression coefficient of SAL is significantly closer to 1 than that of RGBL, indicating SAL prediction is less biased. This observation has been validated by the Bland-Altman plots in **S-Fig. 8(d)**, where SAL produces much smaller bias (9.22) than RGBL (23.32). Except for bias, the width of LOA is also smaller in SAL. When less data was fed, SAL predictions are relatively more stable. However, these two indices of RGBL prediction increased by ~35% (13.08 μmol/L) and ~30% (151.10 μmol/L) with 12.5% (n=40) data feeding, respectively.

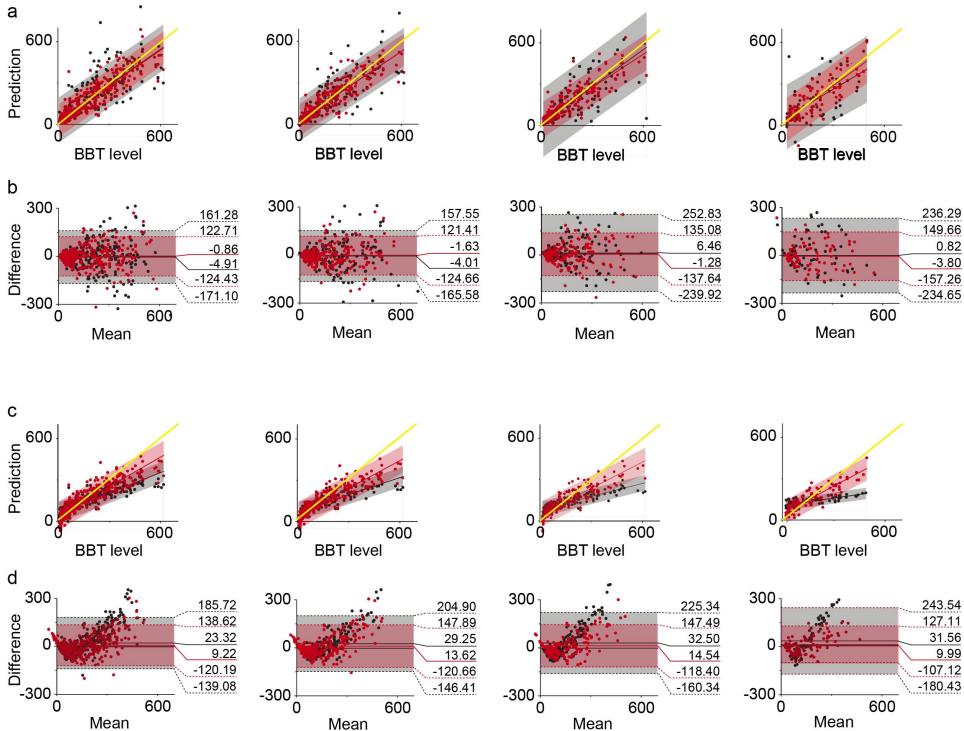

**S-Fig.8. Comparison of prediction performance between SAL and RGBL using ANN and SVM.** a, c, the relationship between BBL and predictions with SAL/RGBL using ANN and SVM, respectively. Black point: RGBL prediction; red point: SAL prediction; black area: 95% prediction band of RGBL prediction; red area: 95% prediction band of SAL prediction. b, d, Bland-Altman plots of predictions with SAL and RGBL using KNN and RF, respectively. Black dotted line: 1.96 limits of agreement of RGBL prediction; red dotted line: 1.96 limits of agreement of SAL prediction; black line: mean error of RGBL prediction; red line: mean error of SAL prediction. All the groups with different sample sizes are randomly selected from the data pool of 320 patients. Units for all values are μmol/L. Resampling percentages from left to right: 100% (n=320); 75% (n=240); 50% (n=160); 25% (n=80).

## S4. Comparison of prediction performance between SAL and RGBL using KNN and RF

Except for ANN and SVM, we also compared the SAL and RGBL predictions using KNN and RF algorithms (**S-Fig. 9**). In summary of these comparisons, SAL improves the prediction quality in varying degrees, especially with less data feeding. To illustrate this point clearer, we quantified the prediction performance of SAL and RGBL with data resampling percentage changing from 12.5% (n=40) to 100% (n=320) with a step width at 6.25% (n=20). The R, MD and STD were then measured and presented as curves in **S-Fig. 9**. The evolution curves showed that the R of SAL always remained in high levels at 0.9 ($p<0.0001$), even when only 12.5% (n=20) of the data was used to train the model. On the contrary, the R of RGBL can be even lower than 0.6. In all methods except for SVM, the prediction biases of SAL are close to 0, smaller or at least comparable to RGBL predictions. In SVM, the bias of SAL is unneglectable, but also 55% (16.87±3.27 μmol/L) smaller than that of RGBL. The STD of MD in SAL slightly increase with smaller sample size but overall lower than 75 μmol/L, which is almost the best level the RGBL can reach.

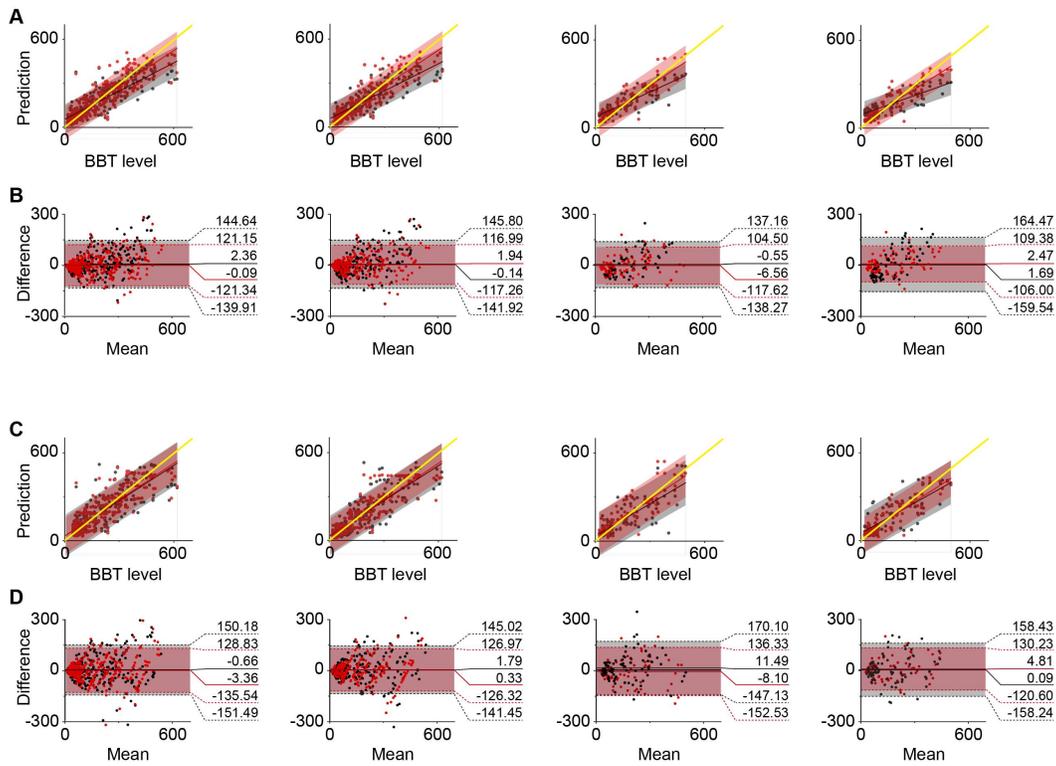

**S-Fig. 9. Comparison of prediction performance between SAL and RGBL using KNN and RF.** a, c, the relationship between BBL and predictions with SAL/RGBL using KNN and RF, respectively. Black point: RGBL prediction; red point: SAL prediction; black area: 95% prediction band of RGBL prediction; red area: 95% prediction band of SAL prediction. b, d, Bland-Altman plots of predictions with SAL and RGBL using KNN and RF, respectively. Black dotted line: 1.96 limits of agreement of RGBL prediction; red dotted line: 1.96 limits of agreement of SAL prediction; black line: mean error of RGBL prediction; red line: mean error of SAL prediction. All the groups with different sample sizes are randomly selected from the data pool of 320 patients. Units for all values are μmol/L. Resampling percentages from left to right: 100% (n=320); 75% (n=240); 50% (n=160); 25% (n=80).

## S5. Stability of predictions in all groups using RGBL and SAL

**S-Fig. 10** shows the bar graphs of prediction qualities in all groups with different sample sizes and algorithms using RGBL and SAL. The SAL owns lower standard deviations of R, MD and STD in all groups than RGBL, implying that SpeCamX-enabled prediction is more stable than conventional methods based on RGB photographs.

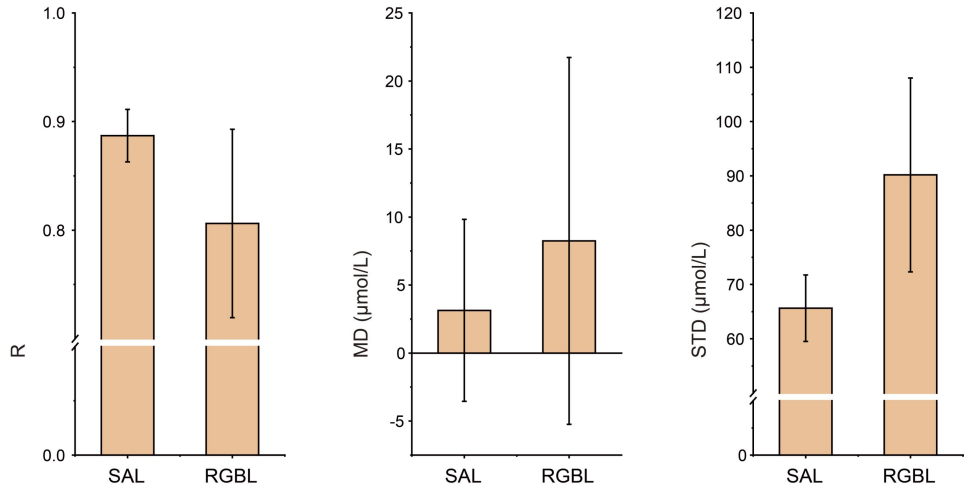

**S-Fig. 10.** Bar graphs of prediction qualities in all groups with different sample sizes and algorithms using RGBL and SAL. SAL-based predictions realized smaller STD than RGBL in R, MD and STD.